# PHENIX Upgrade: Novel Stripixel Detector for Heavy Quark Detection and Proton Spin Structure Measurements at RHIC Energies

Rachid Nouicer*, for the PHENIX Collaboration[1]

Brookhaven National Laboratory, Upton, New York, 11973-5000, U.S.A.

**Abstract:** New design of silicon stripixel sensor has been developed at BNL for PHENIX upgrade. The sensor is a single-sided, DC-coupled, two dimensional position sensitive device with good position resolution. This design is simpler for sensor fabrication and signal processing than the conventional double-sided strip sensor. HPK has produced pre-production stripixel sensors with thickness of 625 μm. The quality assurance tests show that the very low leakage current 0.12 nA per strip allows the use of the SVX4 chip. A long term stability test shows that the leakage current is stable over a long period of time. The study of the effects of irradiation on the performance of the stripixel sensor has been made using p+p collisions at 200 GeV at PHENIX, 14 MeV neutron and 20 MeV proton beams.

*PACS: 25.75.-q; 07.77.-n*

**Keywords:** PHENIX, silicon stripixel sensor, 2d position sensitivity, sensor tests, radiation damage

*Corresponding author: Tel.:+1 6313444342; fax:+1 6313445815; *E-mail address*: rachid.nouicer@bnl.gov

## 1. Motivation for PHENIX detector upgrade

The PHENIX experiment [1] at the Relativistic Heavy Ion Collider (RHIC) at Brookhaven National Laboratory (BNL) has been successful in unlocking the mysteries behind heavy-ion reactions in ultra-relativistic energies [2]. The observables have given important insight into the nature of the matter being produced in Au+Au collisions at RHIC energies but they have also given rise to many intriguing questions. Further work remains however in heavy-ion reactions and also in polarized p+p collisions [2].

The recent results from PHENIX on the suppression [3] and flow [4] of non-photonic electrons are intriguing. However, without an identified sample of charm, the open questions of contribution from semi-leptonic open-beauty decays make a clear interpretation of these results difficult. Efficient topological reconstruction of open charm decays requires a good tracking "point-back" resolution to the primary collision vertex. Further, the beam pipe and innermost layers of detector must be very thin and as close to the beam as possible to allow measurement of particles at low transverse momentum which comprise the bulk of the cross section. A thin beam pipe and inner detector layers are also key elements in efficiently vetoing photon conversion electrons which in combination with electron identification from the PHENIX east-arm spectrometer ($|\eta|<0.35$) will enable much improved measurement of heavy flavor weak decays. Sensitivity of the present measurements of combined bottom and charm quarks will be improved by the central silicon vertex tracker which is proposed as a PHENIX upgrade plan [5] and it will allow us to separate charm and bottom quarks as well as separating them from light flavor quarks.

In the QCD picture, the proton is made from valence quarks, a sea of the quark-antiquark pairs and gluons. The spin of the proton should be explained by the sum of spin of quarks and gluons, and their orbital angular momentum. The contribution from quark spin has been measured by the polarized lepton deep inelastic scattering experiment (DIS). However it is only 20% of the proton spin. Since the gluons can not interact with leptons, DIS is not the best way to investigate gluon contributions. Therefore polarized p+p collisions will use gluons and quarks as probes to interact with gluons. In PHENIX, we have measured the double longitudinal-spin asymmetry $A_{LL}$ of $\pi^0$ production in polarized p+p collisions [6]. In this process, $\pi^o$ carries only a fraction of the momentum of the scattering quark or gluon. More direct information of the gluon polarization will be obtained by measuring $A_{LL}$ of direct-photon and heavy quark production using the PHENIX east-arm spectrometer and the central silicon vertex tracker.

The central silicon vertex tracker consists of four layers of barrel detectors, and covers $2\pi$ azimuthal angle and $|\eta|<1.2$. The inner two layers are silicon pixel detector and the outer two layers are silicon stripixel detectors. Fig.1 illustrates the silicon sub-detectors planned for the upgrade. In this paper, we will discuss the technology choices in the design, the quality assurance tests of the sensors and the study of the effects of irradiation on the stripixel sensors.

## 2. Novel stripixel sensor design and specifications

A novel stripixel silicon sensor has been developed at BNL [7]. The silicon sensor is a single-sided, DC-coupled, two dimensional (2D) sensitive detector. This design is simpler for sensor fabrication and signal processing than the conventional double-sided strip sensor. Each pixel from the stripixel sensor is made from two interleaved implants (a-pixel and b-pixel) in such a way that the charge deposited by ionizing particles can be seen by both implants as presented in Fig1.A. The a-pixels are connected to form a X-strip as is presented in Fig.1.B. The b-pixels are connected in order to form a U-strip as is presented in Fig.1.C. The stereo angle between a X-strip and a U-strip is $4.6^{o}$.

The cross section of the silicon sensor is presented in Fig.3. The basic functionality of the sensor is simple; signal charges (electron-hole pairs) generated for example by particles produced from collisions are separated by the electric field, the electrons moving to the n+ side, holes to the p+ side, thus producing an electric signal which can be amplified and detected. In Fig.3., the first Al layer is the metal contacts for all pixels. All X-strips are routed out by the first metal Al layer. All U-strips are routed out by the second metal Al layer.

The two outer layers (layers 3 and 4) of the central silicon vertex tracker employ silicon stripixel sensors. Each sensor is about 3.43×6.36 cm$^2$. In each long side of the sensor there are 6 sections of bonding pads, with 128 bonding pads each. This implies that each sensor has 2×3×128=768 of X-strips of 80 μm width and 3.1 cm length in beam direction and the same number of U-strips at an angle of 4.6$^o$ to the beam direction. Due to the stereoscopic readout the effective pixel size is 80×1000 μm. Five (for layer 3) or six (for layer 4) sensors are mounted in a ladder. The full length of a ladder in the beam direction is 31.8 cm (for layer 3) or 38.2 cm (for layer 4). A total of 44 ladders are required to cover the azimuth acceptance, 2π. The geometric characteristics of silicon stripixel layers are presented in Table 1.

**Pre-production and quality assurance tests**

Novel stripixel silicon sensor technology developed for the PHENIX upgrade, including the mask design and processing technology, has been transferred from BNL to sensor fabrication company Hamamatsu Photonics (HPK) located in Japan, for mass production. In 2005, HPK delivered the first pre-production wafers with a thickness of 625 μm. Upon request HPK also produced wafers with a thickness of 500 μm by thinning a subset of the 625 μm wafers. The dicing of the pre-production sensors was done at BNL and HPK. A Photo of one stripixel sensor of pre-production is shown in Fig.4.

Stripixel sensor testing of the pre-production has been done at three institutions, BNL, Stony Brook University (SBU), and University of New Mexico (UNM). Each sensor underwent a visual inspection. On each sensor detailed measurements were performed. The 500 μm sensors were found to have a higher leakage current (the guard-ring current: 6 μA at $V_{FD}$=120 V) than 625 μm sensors (the guard-ring current: 300 nA at $V_{FD}$=120 V).

A more detailed test of the current and capacitance as a function of strip were performed at 200 Volts and the results are presented in Fig.5. The results were obtained for one section (128 bonding pads). For given section (128 stripixels), we measured current and capacitance for one stripixel and ground 127 needles by making contact to the pads. The other (11) sections were not grounded. This explains the edge effect observed in the distributions presented in Fig.5 because the current flow from ungrounded stripixels neighbors to the grounded stripixels. These measurements imply that the current and capacitance per strip are low and meet the criteria of silicon sensor acceptance. The terms of acceptance in general of silicon sensor are: 1) visual inspection: good (no scratches), 2) sensor thickness: 625 ± 15 um, 3) silicon resistively: 5-19 kOhm, 4) depletion voltage: 50-250 Volts, 5) junction breakdown > 350 Volts, 6) top of the active area of the sensor has to be passivated except the bonding pads and guard ring and 7) not working stripixels: a) maximum 1% bad stripixel in whole sensor and b) maximum 3 bad stripixels per one section (128 channels).

To measure the total leakage current ($I_{tot}$) of the sensor and eventually to extract the leakage current per strip ($I_{strip} = I_{tot}/N_{tot}$ where the total number of strips $N_{tot}$ = 12 sections x 128 strips = 1536 strips), two sensors that passed the quality assurance tests were mounted on two independent Print Circuit (PCBs) and wire bonded by a private company. The current and capacitance as a function of bias voltage were measured and normalized to $20^o$ C and they are presented in Fig.6. We have studied the stability of the leakage current as function of time for a long period (22 days) and the temperature has been monitored for the same period, see Fig.7. We observe that the leakage current is relatively stable as a function of time and it has good correlation with the temperature of the clean-room.

In Fig.6, the measured total leakage current per stripixel can be obtained as $I_{tot}$ = 179 nA this implies $I_{strip}$ = 179/1536 strips= 0.12 nA which is very low and it allows the use of the SVX4 chips, which have a hard limit of 15 nA/strip. This hard limit comes from that the leakage current from the DC-coupled sensor will rapidly saturate the SVX4 input preamp. The preamp dynamic range is 200 fC, which will saturate in 500 μsec with a 0.4 nA/strip. However, we can issue a preamp-reset signal once per RHIC abort (~ 13 μsec). In fact, this reset frequency limitation puts a hard limit on the maximum acceptable leakage current. If the leakage current stays below this limit, we should not expect problems.

**Radiation damage**

It is well known that the leakage current grows with radiation exposure and also that the increase in leakage current is worse for higher temperature. The radiation study has important implications for the operating temperature requirements for DC-coupled sensors.

To study irradiation effect on the performance of the real silicon stripixel sensor, we mounted two silicon stripixel sensors on two PCBs with all stripxels wire bonded to a single line on each PCB in order to measure the total leakage current before and after irradiation. The two silicon stripixel sensors on PCBs along with 16 silicon diodes were mounted on stand which was installed close to the interaction region (IR) of PHENIX at 10 cm from the beam axis during 2006 Run (p+p collisions). Other stripixel irradiation studies were performed for both 625 and 500 μm thickness as well as silicon diodes using 14 MeV neutron beam at Rikkyo and 20 MeV proton beam at Tsukuba in Japan. The sensors were mounted inside a scattering chamber and Faraday cup to measure the beam current. The radiation exposure is expressed by fluence in neutron-equivalents per $cm^2$ (n-eq/$cm^2$, noted by $\phi_{neq}$). We found that $\phi_{neq.}$ of the beam is consistent with $\phi_{neq}$ of diode which has been evaluated by measuring the increase of the diode leakage current.

The measured increase of leakage current per strip as a function of radiation doses from silicon stripixel sensor using p+p collisions (solid square symbol), 14 MeV neutron beam (open symbol) and 20 MeV proton beam (solid point symbol) are presented in Fig.8. The measured currents were normalized to $20^{o}$ C. In Fig. 8, we also show calculations for an idealized sensor of the expected increase of leakage current versus radiation exposure for different temperature presented by continuous lines. In Fig.8, we observe very nice correlation between the independent measurements, PHENIX-IR, Rikkyo and Tsukuba as well excellent agreement with the calculations for an idealized sensor.

The measurements of radiation dose done by PHENIX during Run 2004 of Au+Au was of the order $2 \times 10^{9}$ n-eq/cm$^2$ in two weeks of running. Assuming scaling of the radiation dose with the collider luminosity, this corresponds to $1 \times 10^{10}$ n-eq./cm$^2$ per week in the future RHIC II environment (x10 increase in luminosity). For 30 weeks of operation per year over a detector design life-span of 10 years, we estimate an expected dose of $3.3 \times 10^{12}$ n-eq./cm$^2$ which is represented by vertical point line in Fig.8. From Fig.8, one can read that in order to keep the current below 15 nA (horizontal dashed line in Fig.8) at a dose $3.3 \times 10^{12}$ n-eq./cm$^2$ we need to be able to operate at low temperature $T < 0^{o}$ C.

 **Summary**

The silicon vertex tracker is an upgrade project for the PHENIX experiment at RHIC. The detector design is comprised of a four-layer barrel, two inner layers of silicon pixel and two outer layers of silicon stripixel with new "spiral" design. In this paper, we present the physics capability added to PHENIX by the new silicon vertex tracker, the technology choices in the design, the tests results of the sensors and the study of the effect of irradiation on the silicon stripixel detector.

References

[1]  K. Adcox et al., Nucl. Instr. and Meth. A499  (2003) 469.

[2]  K. Adcox et al., Nuclear Physics A757  (2005) 184.

[3]  S.S. Adler et al., Phys. Rev. Lett. 96  (2006) 032001.

[4]  S.A. Butsyk et al., Nucl. Phys. A774  (2006) 669.

[5]  http://www.phenix.bnl.gov/phenix/WWW/p/docs/proposals/VTX-PROPOSAL_jul2004.pdf

[6]   S.S. Adler et al., Phys. Rev. Lett. 93  (2004) 202002.

[7]  Li Z. Nucl. Inst. and Meth. A518 (2004) 738.

Figure captions

**Fig.1**: 3D-CAD model from HYTEC of central and forward silicon vertex trackers as well the nosecone calorimeter for PHENIX upgrade.

**Fig.2**. Panel A) stripixel sensor concept with two independent interleaved spiral shaped a-pixel and b-pixel.  Panel B) a-pixels are connected in such way to form a X-strips. Panel C) b-pixels are connected to form a U-strips.

**Fig.3.:** Cross section view of double metal layout of silicon stripixel sensor via contacts on b-pixels of U-strip.

**Fig.4.:** Photo of one silicon stripixel sensor of pre-production produced by HPK.

**Fig.5.:** Panel a) leakage current distribution as a function of strip number of sensor (thickness 625 µm) for one section of bonding pads (128 pads). Panel b) capacitance distribution obtained from the same section discussed in panel a).

**Fig.6.:** Panel a) total leakage current obtained from stripixel sensor presented as a function of bias voltage. Panel b) capacitance distribution obtained in the similar conditions as panel a).

**Fig.7.:** Panel a) long term test of the total leakage current of the sensor (625 µm) as a function of time (22 days). Panel b) measured temperature of the clean-room for the same period as panel a).

**Fig.8.:** Increase in the leakage current per stripixel as a function of radiation dose fluence (in units of 1 MeV neutron equivalents). The continues lines correspond to calculation for idealized stripixel sensor. The horizontal line (dashed line) is 15 nA. The vertical line represents our best estimate for radiation dose after 10 years of operation. The points represent the measurement using real stripixel sensor using p+p collisions in PHENIX-IR (solid point), 14 MeV neutron beam (open symbol) at Rikkyo and 20 MeV proton beam (solid square symbol) at Tsukuba.

| Stripixel | Layer | R3 | R4 |
|---|---|---|---|
| Geometrical dimensions | R (cm) | 10 | 14 |
| | Δz (cm) | 31.8 | 38.2 |
| | Area (cm$^2$) | 1240 | 1600 |
| Channel count | Sensor size R × z (cm$^2$) | 3.43 × 6.36 (384 × 2 strips) | |
| | Channel size | 80 µm × 3 cm (effective 80×1000 µm$^2$) | |
| | Sensors/ladder | 5 | 6 |
| | Ladders | 18 | 26 |
| | Sensors | 90 | 144 |
| | Readout chips | 1080 | 1728 |
| | Readout channels | 138,240 | 221,184 |

**Table 1:** Physical specifications of silicon stripixel layers.

Figure 1.

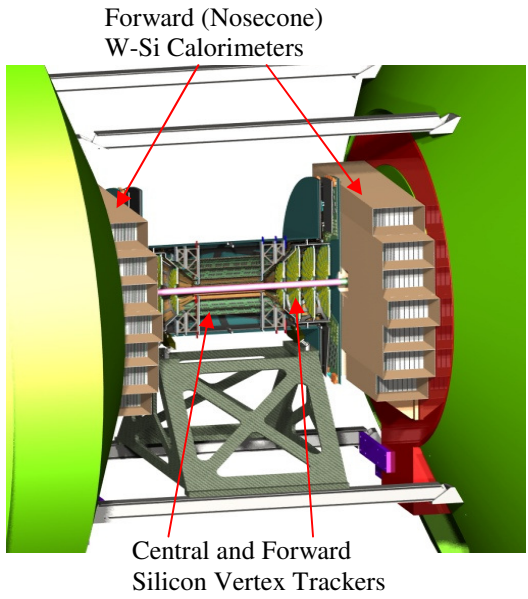

Figure 2.

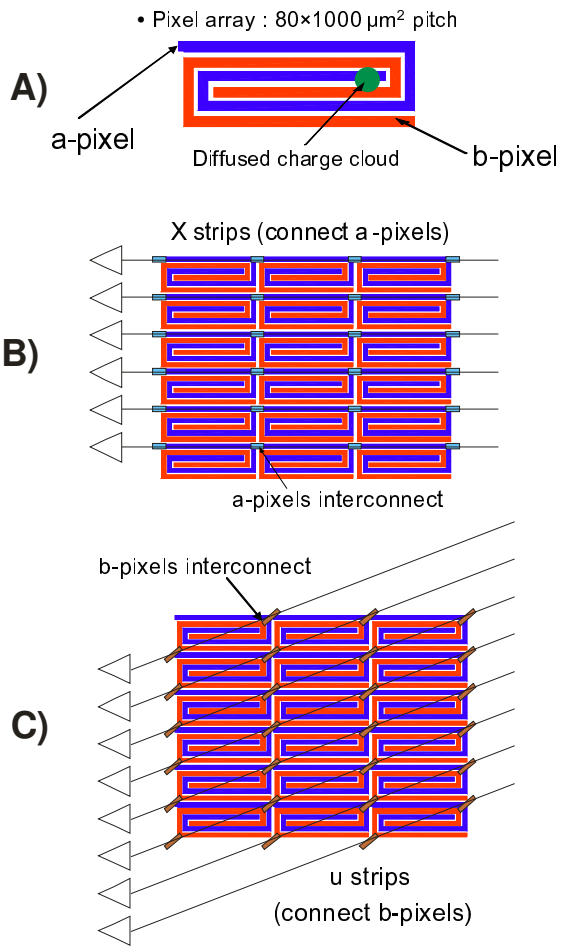

MS199

Figure 3.

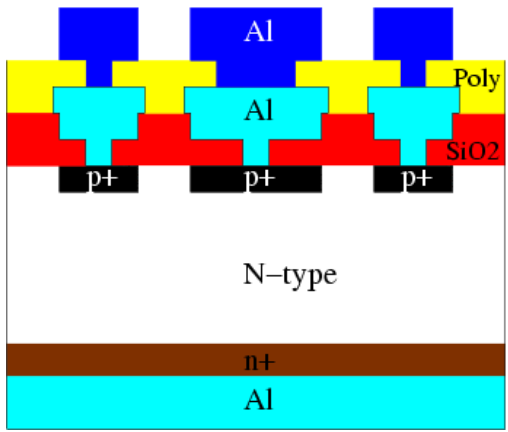

Figure 4.

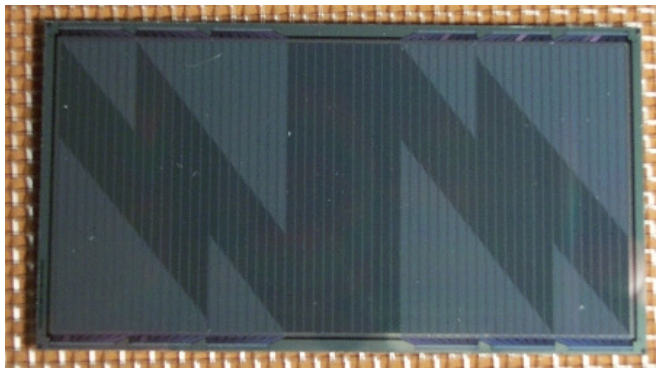

Figure 5.

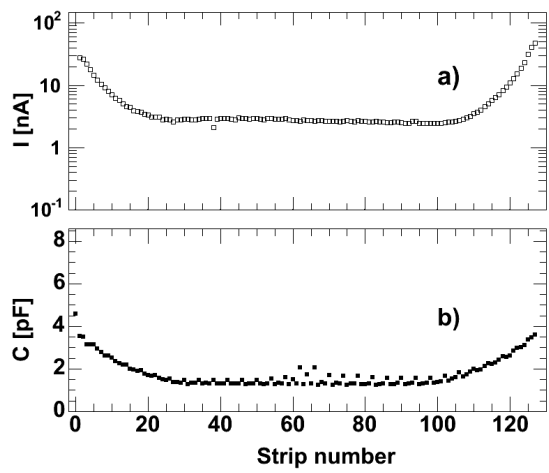

Figure 6.

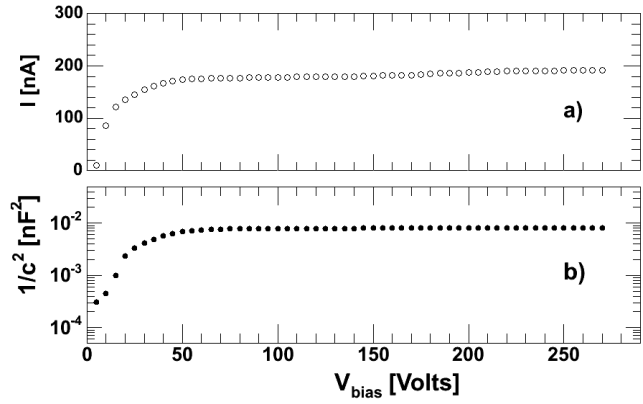

Figure 7.

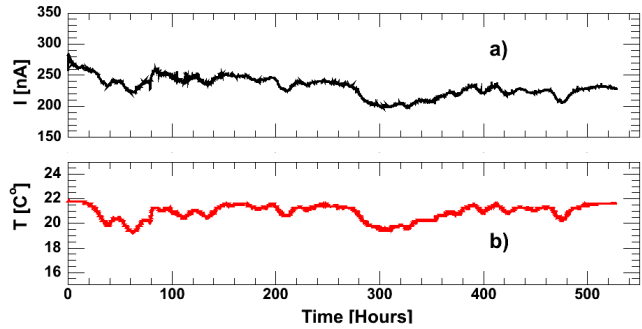

Figure 8.

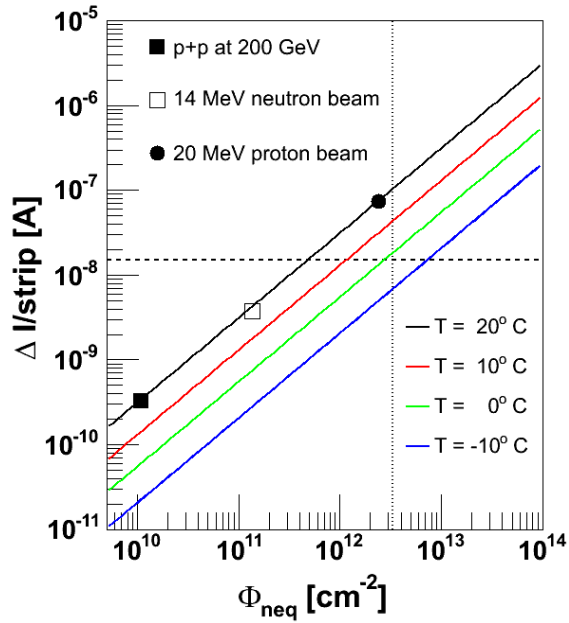